\documentclass[aps,prb,twocolumn,amsmath,amssymb]{revtex4-1} 
\usepackage{graphicx}
\usepackage{tabularx}

\usepackage{nicefrac}
\usepackage{hyperref}
\hypersetup{colorlinks=true, breaklinks=true, citecolor=blue, linkcolor=blue, menucolor=blue, urlcolor=blue}
\usepackage[dvipsnames]{xcolor}
\newcommand{\TeCuChain}{Cu$_2$Te$_2$}
\newcommand{\twosqrt}{$\left(2\sqrt{3} \times \sqrt{3}\right)\textrm{R30}^\circ$}
\newcommand{\twotwosqrtthree}{$\left( 2\sqrt{3} \times 2\sqrt{3}\right)\textrm{R30}^\circ$}
\newcommand{\sqrtthree}{$\left( \sqrt{3} \times \sqrt{3}\right)\textrm{R30}^\circ$}
\newcommand{\rectcell}{$(3\times \sqrt 3 )_\textrm{rect}$}

\begin{document}

\title{CuTe chains on Cu(111) by deposition of 1/3\,ML Te: atomic and electronic structure}



\author{Tilman Ki{\ss}linger}

\author{Andreas Raabgrund}

\author{Begmuhammet Geldiyev}

\author{Maximilian Ammon}
	
\author{Janek Rieger}
	
\author{Jonas Hauner}

\author{Lutz Hammer}

\author{Thomas Fauster} 
	
\author{M. Alexander Schneider}\email{alexander.schneider@fau.de}
\affiliation{Solid State Physics, Friedrich-Alexander-Universit\"{a}t Erlangen-N\"{u}rnberg, Staudtstra{\ss}e 7, 91058 Erlangen, Germany}



\begin{abstract}
 The surface atomic and electronic structure after deposition of \nicefrac{1}{3} monolayer (ML) Te on Cu(111) was determined using a combination of low-energy electron diffraction (LEED), scanning tunneling microscopy and spectroscopy (STM/STS), angle-resolved single and two-photon photoelectron spectroscopy (ARPES /AR-2PPE) and density functional theory (DFT) calculations. Contrary to the current state in literature Te does not create a two-dimensional surface alloy but forms \TeCuChain\ adsorbate chains in a \twosqrt\ superstructure. We establish this by a high-precision LEED-IV structural analysis with Pendry $R$ factor of $R = 0.099$ and corroborating DFT and STM results. The electronic structure of the surface phase is dominated by an anisotropic downward dispersing state at the Fermi energy $E_F$ and a more isotropic upward dispersing unoccupied state at $E-E_F = + 1.43\,\textrm{eV}$. Both states coexist with bulk states of the projected band structure and are therefore surface resonances. 
\end{abstract}


\maketitle



Despite being explored for decades, the structural, thermodynamic and electronic properties of Cu-Te alloys \cite{Pashinkin2005} and derived materials receive on-going interest due to their relevance to back contacts in CdTe/CdS solar cells \cite{Teeter2007,Perrenoud2013,Xia2014,SalGam2018,Du2019} and as thermoelectric materials \cite{He2015} along with tellurides in general \cite{Snyder2008}. 
The latter application is also very promising in a nanoscale form to be incorporated directly into state-of-the-art electronic devices. 
Bulk CuTe has received attention as a material hosting a charge density wave \cite{Zhang2018}.
Tellurides but particularly CuTe nanocrystals have shown distinct plasmonic properties, their dependence on structure has still to be understood \cite{Li2013,Willhammar2017,Agrawal2018, Chen2019}.
Further interest arises as tellurides are important materials in the field of (crystalline) topological insulators, their structural richness and potentially tunable properties open a wide range of applications \cite{Hsieh2012, Zhang2014}.

From all these examples it becomes clear that any analysis of this complex alloy system \cite{Pashinkin2005} must deliver an unequivocal determination of the atomic and electronic structure of the system in question. 
If it comes to thin films or low-dimensional materials, molecular beam epitaxy is a method that allows to exert atomic-scale control over the growing structure and therefore is particularly suited to help revealing structure-property relations.
To understand the nucleation and the formation of interfaces one needs to start with a low Te coverage $\Theta$.
For the particular case of  $\Theta = \nicefrac{1}{3}$ ML of Te on Cu(111) attempts have been made over the years \cite{Andersson1968,Comin1982,King2012, Lahti2014, Tong2020} but a true reliable structural analysis is still missing. 
Seemingly, there is an agreement on a \sqrtthree\ superstructure, but from our work we can infer that such a structure only occurs at much higher Te coverage or under the influence of some contaminant.
In the following we determine the ground-state atomic structure of $\Theta = \nicefrac{1}{3}$ ML of Te on Cu(111) and analyze its electronic structure by a multimethod approach utilizing LEED-IV, STM/STS, ARPES/AR-2PPE and DFT.   


\begin{figure}
	\centering
	\includegraphics[width=0.8\columnwidth]{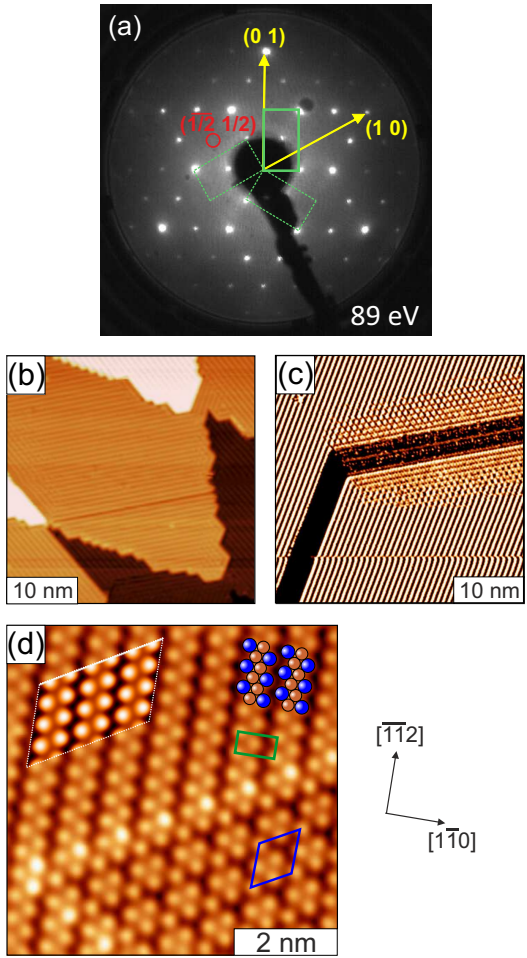}
	\caption{Structural data of \nicefrac{1}{3} ML Te on Cu(111): (a) LEED pattern with the \rectcell\ unit cells of the three possible orientations indicated in green. A glide symmetry plane causes the extinction of e.g. the ($\overline{\nicefrac{1}{2}}\ \nicefrac{1}{2}$) and symmetry related reflexes of each domain. 
   (b) (40 nm)$^2$ image showing the well ordered \rectcell-\TeCuChain\ structure on Cu(111) after annealing at 475\,K. 
   Directions of step edges and \TeCuChain\ chains are coupled.
   (c) After annealing the structure to 870\,K the \rectcell\ structure extends as a single domain over almost a complete terrace and (bunched) step edges cause coherence between terraces. (Derivative image of STM data.) 
   (d) Atomically resolved image of the \rectcell\ structure (upper part) and an exemplary domain boundary between two different orientations of the supercell (lower part, resulting \twotwosqrtthree\ unit cell in blue). 
   The ball model and a DFT STM simulation obtained from the relaxed structure (c.f. Fig.\,\ref{Fig2}) are included for comparison. 
   Imaging parameters: (b) $U = 2.1$\,V, $I = 1.2$\,nA; (c) $U = 1.2$\,V, $I = 0.25$\,nA; (d) $U = -80$\,mV, $I = 1.0$\,nA. (b) and (c) are taken at 300\,K, (d) at 80\,K.	
   }
	\label{Fig1}
\end{figure}

\begin{figure}
	\centering
	\includegraphics[width=0.9\columnwidth]{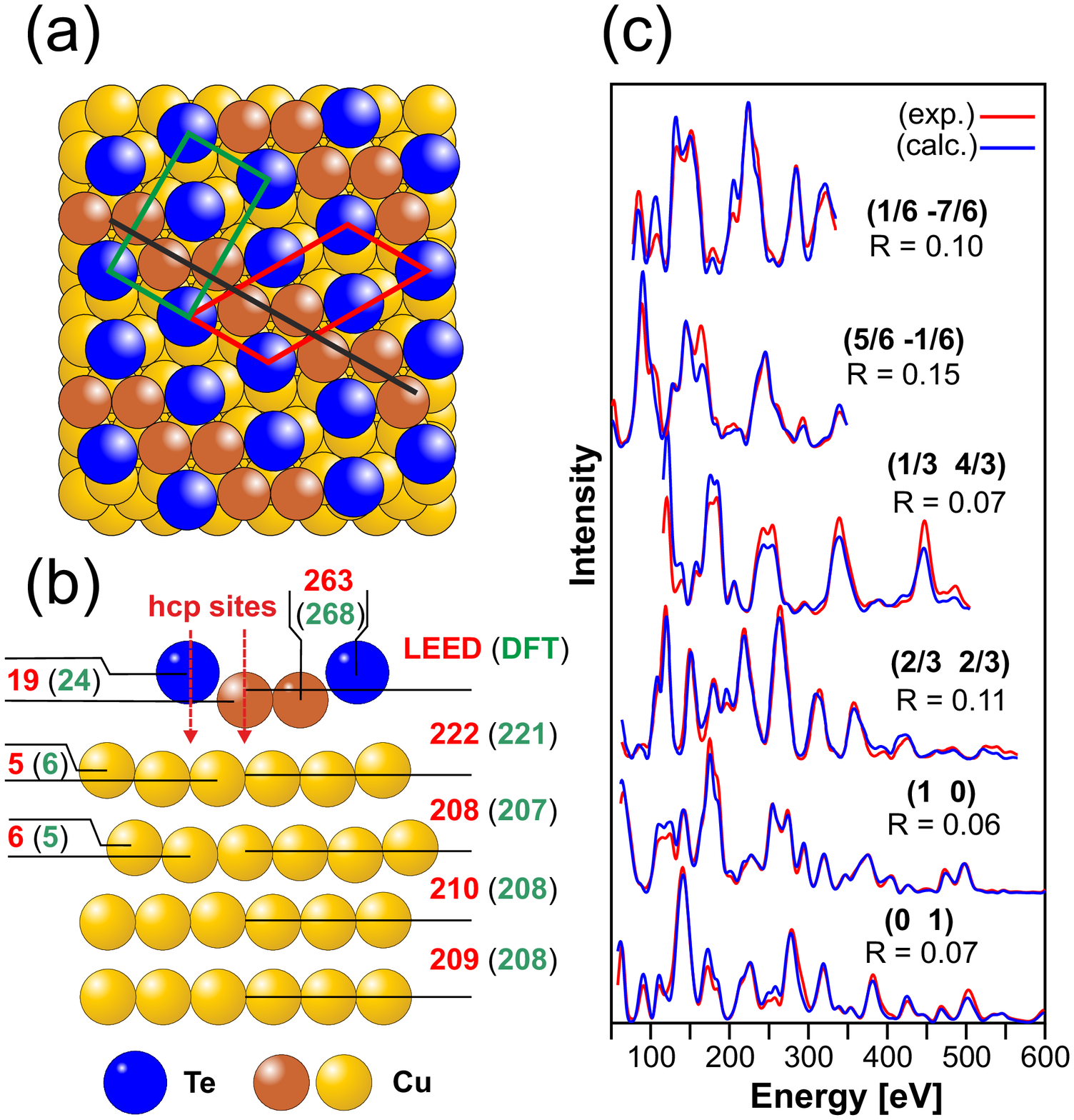}
	\caption{ Resulting surface structure obtained from the LEED-IV analysis with an overall $R$ factor of $R = 0.099$ and corroborated by DFT structural relaxation. (a) Ball model of the structure in top view with the \rectcell\ (green), \twosqrt\ (red) unit cell and glide symmetry plane (black) indicated. (b) Side view with geometrical parameters in pm as determined by LEED (red) and DFT (green). (c) Examples of experimental and calculated LEED-IV spectra of the \rectcell-\TeCuChain\ best-fit structure. 
	}
	\label{Fig2}
\end{figure}

$\nicefrac{1}{3}$ ML of Te was deposited on a clean Cu(111) substrate at 100\,K in ultra-high vacuum (UHV, low $10^{-11}\,\textrm{mbar}$ regime).
Subsequently, the crystal was annealed to temperatures in the range of \mbox{470 to 870\,K}. 
Invariably, this and other variations of the preparation procedure we tested (see \cite{SupMat}) led to the same \twosqrt\ surface structure as identified by LEED patterns (Fig.\,\ref{Fig1}(a)) and LEED-IV spectra (Fig.\,\ref{Fig2}(c)).
In the following we will use an equivalent rectangular cell as primitive unit cell that is denoted as \rectcell\ and indicated in green in Figs.\,\ref{Fig1}(a) and (d).
A characteristic of the LEED pattern is the systematic absence of half-order spots along the main axes indicating a glide symmetry plane of the structure in $[\bar{1}\bar{1}2]$ direction.
The analysis of the LEED patterns revealed that up to annealing temperatures of 570\,K the LEED spots sharpened due to improving lateral order. 
Beyond that temperature the LEED pattern becomes dominated by one or two symmetry related domains of the \rectcell\ superstructure  \cite{SupMat} indicating a tendency to form large single domains on terraces with preferential orientation.

STM imaging reveals the real space topology and
gives a meaningful impression on the crystallographic structure. This can limit the number of possible models that need to be tested in a structural analysis. 
For a proper Te coverage the surface is completely covered by three differently oriented domains of a stripy structure with chains running along [$\bar1\bar1 2$] and equivalent substrate directions (Fig. \ref{Fig1}(b) and (c)). 
With increasing annealing temperature the terraces become more and more mono-domain (Fig. \ref{Fig1}(c) and Supplemental Material (SM) \cite{SupMat}). 
Different to the clean Cu(111) surface, where close-packed [1$\bar10$] step edges dominate, the \rectcell\ phase strongly stabilizes a step orientation along the [2$\bar1\bar1$] chain directions. 
Depending on temperature,
step edges are  rearranged by the superstructure either locally (Fig. \ref{Fig1}(b)) or on a mesoscopic scale (Fig. \ref{Fig1}(c)). 
If the step direction needs to be changed to keep the average surface orientation, narrow stripes of a rotated domain are formed both at the lower and upper terrace of a step, cf. Fig. \ref{Fig1}(c). 
As demonstrated in the SM \cite{SupMat} even an unavoidable miscut of the surface of less than 0.1$^\circ$ leads to a disproportion of domain weights for high temperature annealing. Hence, an \textit{intentional} small misorientation towards [$1\bar10$] may facilitate the preparation of a completely mono-domain surface. 

 
Detailed information on the structure of the \rectcell\ phase comes from atomically resolved images exemplarily shown in Fig.\,\ref{Fig1}(d) that reveal a zig-zag sequence of bright bumps along the chain direction. 
The chains appear separated by troughs of lower height compared to the inner region of the chains themselves. 
The surface density of the bright features is exactly \nicefrac{1}{3} of a monolayer. Consequently, they can at least tentatively be assigned to Te atoms.
At the boundaries between two domains locally a new bridging superstructure with a \twotwosqrtthree\ unit cell is formed to cross-link the two chain directions (c.f. lower part of Fig.\,\ref{Fig1}(d)).
 
For the LEED structural analysis LEED-IV spectra over a total energy range of $\Delta E \approx 8.9\,\mathrm{keV}$ were collected. 
The final best-fit structure reproduced the experimental data with a Pendry $R$ factor of $R = 0.099$ determining in total 35 structural and non-structural parameters with a redundancy factor of $\rho = 15$ (Fig.\,\ref{Fig2}(c)).
The resulting best-fit structure is that of Cu$_2$Te$_2$ chains exhibiting the required glide symmetry plane.
DFT calculations independently prove this structure to be a state of minimal energy. 
The DFT structural relaxation agrees within single picometers with the results from the LEED structural analysis (Fig.\,\ref{Fig2}(b)) and simulated STM images perfectly reproduce the imaging contrast and confirm that the bright features are Te atoms (Fig.\,\ref{Fig1}(d)). 
Both LEED and DFT therefore give a unified picture of the \rectcell\ structure and its impact on the substrate geometry depicted in Fig.\,\ref{Fig2}(a) and (b).
The Cu atoms of the chain reside on hcp  sites of the substrate. 
Te atoms are in total sixfold coordinated to Cu atoms (3 within the chains and 3 to the substrate), while Cu atoms of the chain have in total 8 neighbors, 3 Te and 5 Cu atoms. 
The Te atoms do not sit exactly on hcp sites, but are pushed away by 0.14\,\AA\ perpendicular to the chain direction towards the adjacent bridge site.
As a consequence, first, Te atoms buckle outwards with respect to the Cu plane by 0.19\,\AA\ (DFT: 0.24\,\AA).
Second, within the chain the Te-Cu bonds along [$\bar{1} 0 1$] are with 2.63\,\AA\ significantly shorter than those along [$\bar{1} 1 0$] (LEED: 2.71\,\AA, DFT: 2.69\,\AA). 
And lastly, the Te-Te distance across the troughs is only 4.17\,\AA\ compared to 4.66\,\AA\ across the chain, hence there is no space for additional Cu atoms between the chains.  
Further details of the structural analysis including the best-fit values of all parameters, their errors, and the distortion of the substrate are discussed in the SM \cite{SupMat}.


\begin{figure*}[htb]
	\centering
	\includegraphics[width=0.9\textwidth]{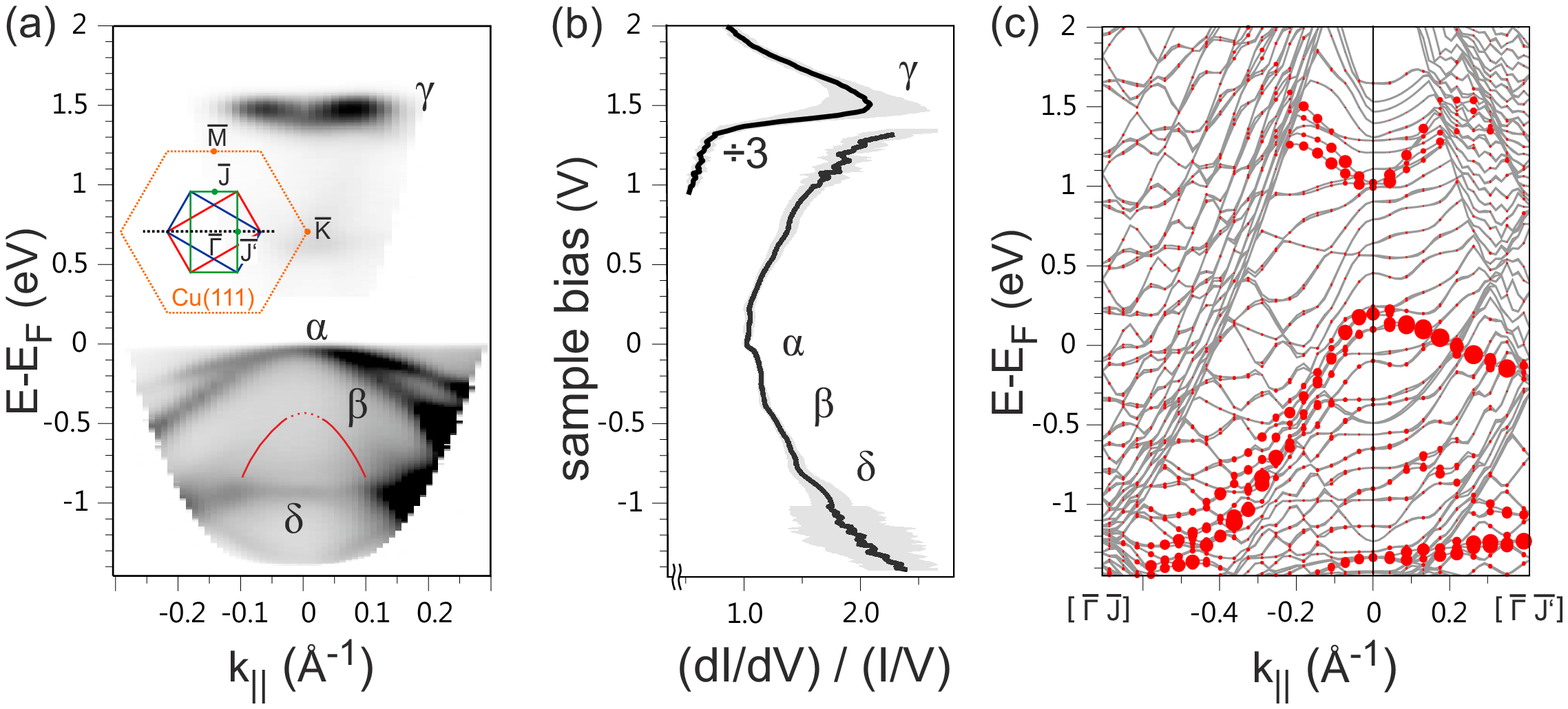}
	\caption{Surface electronic structure of Cu(111)-\TeCuChain-\rectcell. Energy axis aligned across panels. (a) Bottom: laser-based ARPES using \textit{s}-polarized photons with $6.20\,\textrm{eV}$ energy.
		Top: AR-2PPE using \textit{s}-polarized photons with $1.55\,\textrm{eV}$ energy and \textit{p}-polarized photons with $4.65\,\textrm{eV}$. 
		Insert: the data were taken along $[\overline\Gamma\,\overline {\textrm{K}}]$ of the Cu(111) surface Brillouin zone (BZ, black dotted line). Due to domain averaging (green, red and blue BZ of the \rectcell\ cell) the signal contains contributions from states with $k_\parallel$ perpendicular and at $30^\circ$ to the chains.
		The feature $\beta$ is too weak to be presented in the gray scale plot along with features $\alpha$ and $\gamma$. Its analysis is presented in the SM \cite{SupMat}. 
		(b) $\mathrm{d}I/\mathrm{d}V$-STS measurement normalized as $(\mathrm{d}I/\mathrm{d}V)/(I/V)$. The curve represents an spatial average of 380 spectra. The gray band represents the standard deviation of individual spectra. 
		(c) DFT band structure calculation including spin-orbit coupling of a 14 layer Cu(111) slab with the \TeCuChain\ on one side of the slab. The path in the BZ shown is along the chains ($[\overline\Gamma\,\overline {\textrm{J}}]$) and perpendicular to them ($[\overline\Gamma\,\overline {\textrm{J'}}]$).
		Red circles scale with the weight of the states within the \TeCuChain\ and the first Cu layer of the system. 
	} 	\label{Fig3}
\end{figure*}

We investigated the surface electronic structure of the chain phase using ARPES, AR-2PPE, STS, and DFT. 
The results are depicted in Fig.\,\ref{Fig3}.
Applying laser-based ARPES and AR-2PPE we identify four electronic features labeled $\alpha, \beta$, $\gamma$, and $\delta$ in Fig.\,\ref{Fig3}(a). 
The crystal was aligned such that $k_{||}$ is along $\overline{\Gamma}\,\overline{K}$ of the Cu(111) Brillouin zone (BZ). 
Due to the presence of three equivalent rotational domains, the data correspond to a superposition of emission along a $k$-path perpendicular to the chains and at $\pm 30^\circ$ to the chains as indicated in the inset of Fig.\,\ref{Fig3}(a).
Hence, the splitting of the feature $\alpha$ is not attributed to a spin-orbit split state, but to the superposition of a more weakly dispersing band perpendicular to the chains $[\overline\Gamma\,\overline {\textrm{J'}}]$ and a direction 30° off the  chains which has a stronger dispersion similar to that along $[\overline\Gamma\,\overline {\textrm{J}}]$.
A slight intensity around $E-E_F= 0.7\,\textrm{eV}$ is caused by the two-photon excitation of the state $\delta$ (c.f. Fig. S2 of \cite{SupMat}).
The very weak feature $\beta$ is not easily presentable in the grayscale plot without saturating the signal of the other states.
The proof of its existence and the analysis of its dispersion is presented in the SM \cite{SupMat}.

We compare the ARPES data with the spatial average of normalized $\mathrm{d}I/\mathrm{d}V$ STS data (Fig.\,\ref{Fig3}(b)). 
The gray band represents the standard deviation of 380 individual spectra.
We interpret the missing spatial dependence of the spectra as a consequence of the lateral extension of the STM tip that does not allow to probe the "troughs" of the structure separately from the chains.
The features identified by photoelectron spectroscopy show up clearly as onsets ($\alpha, \gamma$) or strong increase in slope ($\beta, \delta$) at the same energies. 


To interpret the experimental data we performed band structure calculations including spin-orbit coupling (Fig.\,\ref{Fig3}(c)). 
We used a relatively large slab consisting of the \TeCuChain\ structure added on 14 layers of Cu(111). 
By this it becomes apparent that the bandgap at the L point of Cu(111) ($\overline\Gamma$) is completely closed by back-folded bulk bands. 
The effect of the spin-orbit coupling is found to be small ($\leq 50\,\textrm{meV}$ ) and only affects states $\alpha$ and $\gamma$.
By analyzing the weight of the electronic states in the first two layers, i.e. including the first complete Cu layer of the substrate, we can identify all experimentally observed features with bands that arise from hybridization of the Te 5p-states with Cu 4sp and 3d states, respectively.
The analysis also shows that the apparent splitting of the $\alpha$-state in the ARPES data of Fig.\,\ref{Fig3}(a) measured along $\overline{\Gamma}\,\overline{K}$ of the Cu(111) BZ is entirely due to the clear anisotropy of the electronic structure along and perpendicular to the \TeCuChain\ chains ($[\overline\Gamma\,\overline {\textrm{J}}]$ and $[\overline\Gamma\,\overline {\textrm{J'}}]$ respectively).
The onset energies of the corresponding states of the three methods are given in Table\,\ref{onsets}.

\begin{table}
	\caption{Energies in eV of the measured and calculated states relative to the Fermi level. $\Phi$: work function, IPS1: $n=1$ image-potential state.}
	\renewcommand{\arraystretch}{1.2}	\setlength{\tabcolsep}{2ex}
	\begin{tabular}{c|r|r|r}
		
		state &ARPES/2PPE& STS& DFT\\	
		\hline 
		$\delta$	&$-0.89\pm0.05$  &$-0.78\pm0.06$  & $-1.3$ \\
		$\beta$	&$-0.43\pm0.05$  &$-0.38\pm0.05$  & $-0.4$ \\
		$\alpha$	& $-0.06\pm0.05$ &$0.01\pm0.05$ & $0.2$ \\ 
		$\gamma$	& $1.43\pm0.05$ &$1.35\pm0.08$ & $1.0$ \\ 
		$\Phi$ & $4.85\pm0.05$ & --- & --- \\
		IPS1	& $4.25\pm0.05$ & --- & --- 
	\end{tabular} 
	\label{onsets}
\end{table}


Our results contradict the findings of others who claim that the deposition of \nicefrac{1}{3} ML Te leads to the formation of a Cu$_2$Te-surface alloy with a \sqrtthree\ supercell \cite{King2012,Lahti2014,Tong2020}.
Only one early investigation reported a \twosqrt\ structure for this coverage \cite{Comin1982}. 
Using surface extended x-ray absorption fine-structure it was found that the structure consists of Te-Cu bonds oriented predominantly within the surface plane compatible with our findings. 
A more detailed model could not be provided \cite{Comin1982}. 
Although none of the above papers could provide structural data at the level discussed here, we investigated this matter thoroughly by checking  independently the preparation and evolution of phases when depositing Te in all three UHV chambers used in our study.
Only in one of our UHV systems we observed the conversion of the \rectcell\ to a \sqrtthree\ structure on two different Cu(111) crystals after prolonged exposure to the running LEED optics at room temperature. 
The \rectcell-phase could be fully recovered by mild annealing of the sample to about 450\,K proving that the \sqrtthree\ structure is metastable.
This behavior is clearly different to the stable AgTe \sqrtthree\ honeycomb on Ag(111) investigated with the same instruments \cite{Uenzelmann2020}.
From our experiments it became clear that the transition involves some unknown adsorbate emitted by the electron source of the LEED optics. 
However, what exactly triggered the transformation remained unclear, adsorption studies using H$_2$, CO, CO$_2$, and O$_2$ could exclude these agents.

To establish the relation between various \sqrtthree superstructures and the \TeCuChain\ chain  structure presented here, we calculated the phase formation energy $\Delta E_f$ of all phases within the same supercell.
\begin{equation*}
\Delta E_f = E_{\sigma} - E_{\textrm{CuTe-chain}} \pm n E_{\textrm{Cu}} 
\label{eq1}
\end{equation*}
where $E_{\sigma}$ denotes the calculated total energy of the particular phase $\sigma$, $E_{\textrm{CuTe-chain}}$ is the total energy of the \TeCuChain-\rectcell\ phase, and the last term accounts for the total (bulk) energy $E_{\textrm{Cu}}$ of $n$ additional Cu atom(s) that make up the phase $\sigma$. 
The \TeCuChain-\rectcell\ phase with top-layer Cu atoms in hcp hollow sites has the lowest energy and therefore is the ground state of the system. 
A shift in registry to an fcc stacking which is incompatible to our LEED data causes an increase in total energy of $\Delta E_f = 30\,\textrm{meV}$.
We find that the honeycomb phase in hcp stacking which only differs from the chain phase by a hop of one of the two Cu atoms to an adjacent hcp hollow site also has $\Delta E_f = 30\,\textrm{meV}$. 
In contrast, the surface alloy (\rectcell-Te$_2$Cu$_4$) comes at the expense of $\Delta E_f = 370\,\textrm{meV}$.
Hence, the occurrence of the Cu(111)-\sqrtthree-Cu$_2$Te can be safely excluded. 
The calculations indicate that the occurrence of the mentioned metastable \sqrtthree\ phase triggered by additional adsorbates may be a transformation of the \TeCuChain\ ground-state phase to the honeycomb phase. 
Therefore, we suggest that in Refs. \cite{Lahti2014,Tong2020} this phase was investigated but interpreted in the framework of a surface alloy. 
The ARPES data shown in Ref.\,\cite{Tong2020} would support this view \cite{TongComment}.

A recent paper investigated the growth of Te on Cu$_2$Sb(111) under very similar preparation conditions as the present system \cite{Zhou2020}. 
The authors claim that Te forms zigzag chains on the substrate that have a significant distance of 2.99\,\AA\ from the surface and are therefore decoupled. 
Although the Sb from the CuSb(111) substrate might introduce a different Te growth behavior than on Cu(111), we would like to point out that the DFT simulation derived from our \TeCuChain\ chain model (\ref{Fig1}(d)) fits much better the STM images of phase S2 than simulations based on Te$_2$ chains (Figs.\,1 and 2 of Ref.\,\cite{Zhou2020}) suggesting that the two have the same structure.

In conclusion, by a LEED-IV structural analysis with Pendry $R$ factor $R = 0.099$ we have established that the ground state of \nicefrac{1}{3} ML Te on Cu(111) is a chain-like \twosqrt\ or equivalently \rectcell\ superstructure in which \TeCuChain\ chains are formed on top of the otherwise unreconstructed surface. 
The Te and Cu atoms in the chains sit on or close to hcp hollow sites.
DFT calculations confirm all structural parameters of the LEED analysis independently.
At the correct Te coverage and after annealing steps the structure exhibits a close to perfect order with very little point defects and domain boundaries. A mutual influence between surface steps and the orientation of the \TeCuChain\ chains was observed.
The analysis of the electronic structure by ARPES, AR-2PPE, and STS reveals two dominant Te induced states on the surface: a downward dispersing and anisotropic occupied state at the Fermi energy $E_F$ and an upward dispersing unoccupied state at $E-E_F = 1.43$\,eV.
DFT calculations indicate a spin-orbit splitting of the order of 50\,meV.
Thus we arrive at a complete structural and electronic understanding of the system.

Previous attempts to unravel the surface structure resulted in inconsistent findings of experiment and theory. 
The difficulties arise due to the presence of a metastable CuTe honeycomb phase in a \sqrtthree\ superstructure that has been interpreted as a substitutional alloy.  
Our findings clearly disprove the occurrence of a substitutional Cu$_2$Te alloy and reveal the true initial growth mode also relevant for the growth of thicker CuTe films on Cu(111).

\vspace{5mm}
M.A.S. gratefully acknowledges supply of computing resources and support provided by the Erlangen Regional Computing Center (RRZE).

\bibliography{literatur_TeCu111-2sqrt3xsqrt3}

%
%
%
%
%

\end{document}